\documentclass[a4paper,showpacs,showkeys,amsmath,amssymb,superscriptaddress,12pt,tightenlines]{revtex4}
\begin{document}
\title{Rastall's gravity equations and Mach's Principle}
\author{Vladim\'{\i}r Majern\'{\i}k}
\email{majere@prfnw.upol.cz}
\affiliation{Department of Theoretical Physics, Palack\'y University, 17.
  listopadu 50, Olomouc, CZ-772\,07, Czech Republic}
\affiliation{Institute of Mathematics, Slovak Academy of Sciences,
\v{S}tef\'anikova 19, Bratislava, SK-814\,73, Slovak Republic}
\author{Luk\'a\v{s} Richterek}
\email{richter@prfnw.upol.cz}
\affiliation{Department of Theoretical Physics, Palack\'y University, 17.
  listopadu 50, Olomouc, CZ-772\,07, Czech Republic}
\date{\today}%

\begin{abstract}
  Rastall~\cite{rastall:1972} generalized Einstein's field equations
  relaxing the Einstein's assumption that the covariant divergence of the
  energy-momentum tensor should vanish. His field equations contain a free
  parameter $\alpha$ and in an empty space, i.e. if $T_{\mu\nu}=0$, they
  reduce to the Einstein's equations of standard general relativity.  We
  calculate the elements of the metric tensor given by Rastall' equations
  for different $\alpha$ assuming that $T_{\mu\nu}$ to be that of a perfect
  fluid and analyse these model solutions from the point of view of Mach's
  principle.  Since the source terms in Rastall's modified gravity
  equations include the common energy-momentum tensor $T_{\mu\nu}$ as well
  as the expressions of the form $\left(1-\alpha\right)g_{\mu\nu}T/2$, these
  source terms depend on metric determined by the mass and momentum
  distribution of the external space. Likewise, in the classical limit, the
  source term of the corresponding Poisson equation for $\alpha\neq 1$
  depends on the gravitational potential in the sense of Mach's principle.
\end{abstract}

\maketitle

\section{Introduction}
\label{sec:intro}

General relativity assumes that particle rest masses are space-time
constants independent of the space-time metric in which they occur. The
experimental evidence relating to this assumption is fragmentary and far
from establishing it on a firm basis.  The space-time constancy of particle
mass is a typical feature also of Newton's mechanics.  Mach in his critique
of the Newtonian mechanics~\cite{mach:1933} pointed out to the fact that
mass and momentum distributions of the external space should be linked with
inertia of a mass body.  This approach to treat problems of mechanics is
commonly called as Mach's principle~\cite{einstein:1956}.  Many versions of
Mach's principle can be found in the literature (see, e.g.
\cite{dicke:1964,einstein:1956,nightingale:1977}) and unique, satisfactory
``canonical'' formulation of the principle does not seem to exist as yet.
Nevertheless, there seem to be a fair consensus about the consequences
which should follow from it.  Mach's principle requires that the presence
and motions of large external masses in the universe must have some
definite effect upon the mass objects occurring in it.  As is well-known,
Mach's principle is not incorporated in the Einstein equations of classical
general relativity. To be included, the standard general relativity
equations have to be modified.  In the literature, there are several
attempts to modify the Einstein equations in such a way that they
incorporate at least some requirements following from Mach's principle
(see, e.g.~\cite{weinberg:1972}).

One of the fundamental assumptions of the Einstein theory of gravitation is
vanishing divergence of the energy-momentum tensor
\begin{equation}
  \label{eq:conscond}
  \nabla_{\nu}T^{\mu\nu}=0.
\end{equation}
However, as stressed by Rastall~\cite{rastall:1972} and others (see
e.g.~\cite{smalley:1975}), the local conservation of the energy-momentum
tensor expressed by this relation has not been specifically tested by
observation.  This is why Rastall~\cite{rastall:1972} generalized the
Einstein field equations by relaxing this assumption.  Later, Al-Rawaf and
Taha~\cite{alrawaf:1996grg} found an alternative, equivalent form of
Rastall's equations introducing a free parameter $\alpha$ and their
modified equations used throughout this paper reduce to the Einstein ones
when setting $\alpha$ equal to $1$. Rastall's equations reduce to those of general
relativity for {\it empty} space independently of the value
of the parameter $\alpha$. Thus, all the crucial tests of general
relativity (the perihelion advance of planets in our solar system, the
deflection of light, the gravitational red shift and the delay of radar
echos) remain valid also in Rastall's theory.

Rastall's modified equations becomes important mainly when treating general
relativistic problems including sources.  Since the source terms in Rastall's
modified gravity equations include the components of the common
energy-momentum tensor $T_{\mu\nu}$ as well as additional terms of the form
$\left(1-\alpha\right)g_{\mu\nu}T/2$, the source terms of Rastall's
equations depend on the metric determined by the mass and momentum
distributions of the external space. This is consistent with an important
requirement following from Mach's principle \cite{nightingale:1977}.

In the Newtonian limit for $\alpha \neq 1$, Rastall's equations with source
terms lead to a modified Poisson equation whose source depends on the
gravitational potential, which also puts a strong constraint on $\alpha$.
It should be approximately equal to 1 in order to provide the Poisson
equation in common weak gravitational fields. However, in large mass
concentration and in strong gravitational field, i.e.  in the stellar
dynamics, even this small deviation of $\alpha$ from $1$ might play a
significant role.

In what follows we try to show that solutions of Rastall's equations for
some model gravitational systems satisfy at least one requirement following
from Mach's principle, namely that the source terms in field equations
depend, either on metric elements in case of general relativistic problems
or on the gravitational potential in the classical Newtonian case.  In
this sense they seem to be more 'Machian' than their standard general
relativistic counterparts.

The article is organized as follows. In Section~\ref{sec:rasteq} we briefly
summarize Rastall's theory rewriting his equations in the form in which the
source terms are written as sums of ordinary energy-momentum tensor
$T_{\mu\nu}$ and an additional terms of the form
$\left(1-\alpha\right)g_{\mu\nu}T/2$.
$T=T^{\mu}_{\hphantom{\mu}\mu}$, $\alpha$ being a free parameter.  In
Section~\ref{sec:sphersym} we solve Rastall's equations for the
isotropic perfect fluid calculating the components of metric tensor for
arbitrary $\alpha$. In Section~\ref{sec:mach} we discuss the
Machian property of Rastall's equations pointing out to the fact that in
the classical ``Newtonian'' limit the source term (i.e. mass density)
necessarily depends on the gravitational potential provided $\alpha\neq 1$.

Throughout the text geometrized units in which $c=1$, $G=1$ are used. The metric
signature $-+++$, notation and sign conventions follow~\cite{carroll:2003,hartle:2003}.

\section{Rastall's gravity equations}
\label{sec:rasteq}

The standard Einstein field equations read as (see, e.g.~\cite{hartle:2003})
\begin{equation}
  \label{eq:eeq}
G_{\mu\nu}=R_{\mu\nu}-\frac{1}{2}\,g_{\mu\nu}R=8\pi T_{\mu\nu},
\end{equation}
where $R=R^{\mu}_{\hphantom{\mu}\mu}$; an alternative form of Eq.~(\ref{eq:eeq}) is
\begin{equation}
  \label{eq:eeq2}
  R_{\mu\nu}=8\pi\left(T_{\mu\nu}-\frac{1}{2}\,g_{\mu\nu}T\right).
\end{equation}
As it was demonstrated by Al-Rawaf and Taha~\cite{alrawaf:1996grg}, the
Rastall's modification of Einstein theory can be obtained by substituting a
more general form of the Einstein tensor
$${\cal G}_{\mu\nu}=\alpha R_{\mu\nu}+\beta g_{\mu\nu}R$$
for the left-hand side of the Eq.~(\ref{eq:eeq}), where $\alpha$ and
$\beta$ are constants. In standard general relativity
the corresponding values $\alpha=1$ and $\beta=1/2$ are
determined by the requirements that~\cite{weinberg:1972}
\begin{itemize}
\item[(i)] Eq.~(\ref{eq:eeq}) becomes the classical Poisson equation
for gravitational potential in the Newtonian limit and that
\item[(ii)] energy-momentum is locally conserved, i.e.
  Eq.~(\ref{eq:conscond}) holds.
\end{itemize}

However, if we relax the latter condition, then the constants $\alpha$ and $\beta$ are linked
only through the relation~\cite{alrawaf:1996grg}
\begin{equation}
  \label{eq:alfbet}
  \beta=\frac{\alpha\left(\alpha-2\right)}{2\left(3-2\alpha\right)},\qquad \alpha\neq 0,\ \frac{3}{2}.
\end{equation}
Using Eq.~(\ref{eq:alfbet}) one obtains the generalized field equations
first published by Rastall~\cite{rastall:1972}
\begin{equation}
  \label{eq:raseq}
  R_{\mu\nu} -\frac{1}{2}\,\gamma g_{\mu\nu} R=kT_{\mu\nu},
\end{equation}
where $\gamma$ and $k$ are the following functions of the parameter $\alpha$
$$\gamma=\frac{2-\alpha}{3-2\alpha},\qquad k=\frac{8\pi}{\alpha}.$$
Evidently, if $\alpha=1$ (i.e. also $\gamma=1$), then Eq.~(\ref{eq:raseq}) reduces
to Eq.~(\ref{eq:eeq}), so the Rastall's theory includes standard general
relativity as its special case.

The conservation condition~(\ref{eq:conscond}) is not generally satisfied
in Rastall's theory and according to Eq.~(\ref{eq:eeq2}) it is
replaced by~\cite{alrawaf:1996grg}
$$\nabla_{\nu}T^{\nu}_{\mu}=-\frac{1-\alpha}{2k\left(3-2\alpha\right)}\,\partial_{\mu}R=
-\frac{1-\alpha}{2}\,\partial_{\mu}T.$$
To make use of already known solutions of general relativity it is
convenient to rewrite Eq.~(\ref{eq:raseq}) into the conventional form employing the standard
Einstein tensor
\begin{equation}
  \label{eq:rasteq}
G_{\mu\nu}=R_{\mu\nu}-\frac{1}{2}\,g_{\mu\nu}R=k{\cal T}_{\mu\nu}=\frac{8\pi}{\alpha}\,{\cal T}_{\mu\nu},
\end{equation}
where
\begin{equation}
  \label{eq:emrast}
{\cal T}_{\mu\nu}=T_{\mu\nu}+\frac{1-\alpha}{2}\,g_{\mu\nu}T.
\end{equation}
Therefore, it is formally possible to solve Eq.~(\ref{eq:rasteq}) like the
standard Einstein equations~(\ref{eq:eeq}) with the modified energy-momentum
tensor ${\cal T}_{\mu\nu}$ which itself need not correspond to any physically
realistic situation. For example, for the perfect fluid, the model important
both in astrophysics and cosmology, the energy momentum tensor reads as
\begin{equation}
  \label{eq:perff}
  T_{\mu\nu}=\left(\varrho+p\right)u_{\mu}u_{\nu}+pg_{\mu\nu},
\end{equation}
with the density $\varrho=\varrho(t)$, pressure $p=p(t)$ and four-velocity $u$
fulfilling the condition $u_{\mu}u^{\mu}=-1$. Using
Eq.~(\ref{eq:emrast}) and the fact that $T=3p-\varrho$, we can introduce the tensor
\begin{equation}
  \label{eq:emover}
  {\cal T}_{\mu\nu}=\left({\cal R}+{\cal P}\right)u_{\mu}u_{\nu}+{\cal P}g_{\mu\nu},
\end{equation}
where
\begin{gather}
  \label{eq:rhoover}
  {\cal R}=\frac{1}{2}\left(3-\alpha\right)\varrho-\frac{3}{2}\left(1-\alpha\right)p,\\
  \label{eq:pover}
  {\cal P}=-\frac{1}{2}\left(1-\alpha\right)\varrho+\frac{1}{2}\left(5-3\alpha\right)p.
\end{gather}
The linear dependence of the functions ${\cal R}$ and ${\cal P}$ on
$\varrho$, $p$ for some values of $\alpha$ is given in Table~\ref{tab:alphpr}; it
will be demonstrated below that the value $\alpha=3$ is not acceptable for
our observed physical universe.
\begin{table}
  \centering
  \begin{tabular}{|c|c|c|}
\hline
$\alpha$ & ${\cal{P}}(r,\alpha)$ & ${\cal{R}}(r,\alpha)$\\
\hline
$-1$& $4p-\varrho$&$-3p+2\varrho$\\
$1/2$& $7p/4-\varrho/4$&$-3p/4+5\varrho/4$\\
$1$ &$p$&$\varrho$\\
2&$-p/2+\varrho$&$3p/2+\varrho/2$\\
$3$&$-2p+\varrho$&$3p$\\
\hline
\end{tabular}
  \caption{The functions ${\cal R}$ and ${\cal P}$ for various values of
    the parameter $\alpha$; let us remind that the values $\alpha=0,3/2$ are
    excluded by Eq.~(\ref{eq:alfbet})}
  \label{tab:alphpr}
\end{table}
Moreover, the Bianchi identities and Eq.~(\ref{eq:rasteq}) ensure the
vanishing divergence of introduced tensor ${\cal T}^{\mu\nu}$, i.e.
\begin{equation}
  \label{eq:diverg}
  \nabla_{\nu}{\cal T}^{\mu\nu}=0.
\end{equation}
Therefore perfect fluid solutions of Rastall's equations~(\ref{eq:raseq})
or (\ref{eq:rasteq}) are formally equivalent to the solutions of Einstein
equations~(\ref{eq:eeq}) for the fluid with ``density''
${\cal R}$ and ``pressure'' ${\cal P}$ given
by~(\ref{eq:rhoover}) and~(\ref{eq:pover}) respectively. We exploit this
fact when studying particular cases.

In the following sections we assume that $\alpha$ is a universal parameter
with the same value for all physical systems. It is always important to
check that obtained results reduce to known solutions of standard general
relativity for $\alpha=1$.

\section{Static spherical symmetric case}
\label{sec:sphersym}

First, we find the static isotropic metric fulfilling the Rastall's
equations~(\ref{eq:rasteq}) assuming that the energy-momentum tensor
$T_{\mu\nu}$ corresponds to the perfect fluid described by
Eq.~(\ref{eq:perff}). Formally, we will follow calculations analogical to
those presented e.g. in~\cite{hartle:2003,schutz:1985}.

The metric of a static, spherically symmetric spacetime can be written in
the form
\begin{equation}
  \label{eq:statmet}
  {\rm d}s^2=-\exp\left[2B\left(r,\alpha\right)\right]\,{\rm
    d}t^2+\exp\left[2A\left(r,\alpha\right)\right]\,{\rm d}r^2+r^2\left({\rm
      d}\vartheta^2+\sin^2\vartheta{\rm d}\varphi^2\right),
\end{equation}
where we introduce two functions $A\left(r,\alpha\right)$ a
$B\left(r,\alpha\right)$ and employ standard spherical coordinates of a
distant observer (we implicitly suppose the asymptotic flatness of the
spacetime). Then, the non-null components of the standard Einstein tensor
read as
\begin{eqnarray}
  \label{eq:etsfa}
  G_{tt}&=&\frac{\exp\left(2B\right)}{r^2}\,\frac{{\rm
      d}}{{\rm
      d}r}\left\{\vphantom{\frac{a}{b}}r\left[1-\exp\left(-2A\right)\right]\right\},\\
\label{eq:etsfb}
  G_{rr}&=&-\frac{\exp\left(2A\right)}{r^2}\,\left[1-\exp\left(-2A\right)\right]+\frac{2}{r}\,\frac{{\rm d}B}{{\rm d}r},\\
\label{eq:etsfc}
  G_{\vartheta\vartheta}&=&
   r^2\exp\left(-2A\right)\,\left[\frac{{\rm d}^2B}{{\rm
         d}r^2}+\left(\frac{{\rm d}B}{{\rm
           d}r}\right)^2+\frac{1}{r}\,\frac{{\rm d}B}{{\rm d}r}
   -\frac{1}{r}\,\frac{{\rm d}A}{{\rm d}r}-\frac{{\rm d}A}{{\rm d}r}\,\frac{{\rm d}B}{{\rm d}r}\right],\\
\label{eq:etsfd}
  G_{\varphi\varphi}&=&\sin^2\vartheta\,G_{\vartheta\vartheta}.
\end{eqnarray}
For a static fluid it also holds $u^{t}=\exp(-B)$, $u_{t}=\exp(B)$ and
$u^{r}=u^{\vartheta}=u^{\varphi}=0$, therefore according to
Eq.~(\ref{eq:perff}) we obtain non-vanishing components of energy-momentum
tensor
\begin{equation}
  \label{eq:emperf}
  T_{tt}=\varrho\exp\left(2B\right),\quad
  T_{rr}=p\exp\left(2A\right),\quad
  T_{\vartheta\vartheta}=pr^{2},\quad
  T_{\varphi\varphi}=\sin^2\vartheta\,T_{\vartheta\vartheta}
\end{equation}
which can be substituted into Rastall's equations~(\ref{eq:rasteq}).

Is is convenient to solve the $(t,t)$ component of these equations through
introducing a different unknown function ${\cal M}\left(r\right)$
defined as
\begin{equation}
  \label{eq:mfce}
  {\cal M}\left(r\right)=\frac{r}{2}\left[1-\exp\left(-2A\right)\right]
\end{equation}
or compared with the metric~(\ref{eq:statmet})
\begin{equation}
  \label{eq:mfceb}
  g_{rr}=\exp\left(2A\right)=\left(1-\frac{2{\cal M}}{r}\right)^{-1}.
\end{equation}
Then the $(t,t)$ Rastall equation implies
\begin{equation}
  \label{eq:rqtt}
  \frac{{\rm d}{\cal M}}{{\rm d}r}=\frac{k}{2}\,r^2{\cal R}=
\frac{4\pi}{\alpha}\,r^2{\cal R};
\end{equation}
substituting for ${\cal R}$ and integrating with the condition ${\cal
  M}\left(0\right)=0$, one gets
\begin{equation}
  \label{eq:rqttint}
  {\cal M}\left(R\right)=\int\limits_{0}^{R}\frac{4\pi}{\alpha}\,r^2{\cal R}\,{\rm d}r=
  \int\limits_{0}^{R}\frac{4\pi}{\alpha}\,r^2
  \left[\frac{1}{2}\left(3-\alpha\right)\varrho-
        \frac{3}{2}\left(1-\alpha\right)p\right]\,{\rm d}r.
\end{equation}
Evidently, setting $\alpha=1$ we come to the familiar result of general
relativity
\[
{\cal M}\left(R\right)=m\left(R\right)=\int\limits_{0}^{R}4\pi\varrho\,r^2\,{\rm d}r,
\]
where $m(R)$ is often called as mass function and outside the fluid,
i.e. for a distant observer, it determines the mass of the studied spherical
symmetric body with a surface at $r=R$.

Using this result we can transform the $(r,r)$ component of
Eq.~(\ref{eq:rasteq}) into the form
\begin{equation}
  \label{eq:rqrr}
  \frac{{\rm d}B}{{\rm
      d}r}=\frac{{\cal M}+4\pi{\cal P}r^{3}/\alpha}{r\left(r-2{\cal M}\right)}.
\end{equation}
As we can see, the two remaining Rastall's equations for
$(\vartheta,\vartheta$ and $(\varphi,\varphi)$ components are completely
equivalent. Instead of solving them explicitly, commonly another equation
is employed instead; substituting from Eq.~(\ref{eq:emperf}) into formal
``conservation law'' (\ref{eq:diverg}) one gets
\begin{equation}
  \label{eq:motion}
  \left({\cal R}+{\cal P}\right)\,\frac{{\rm d}B}{{\rm d}r}=
  -\frac{{\rm d}{\cal P}}{{\rm d}r},\ \mbox{i.e.}
  \left(\varrho+p\right)\,\frac{{\rm d}B}{{\rm d}r}=
  -\frac{{\rm d}}{{\rm d}r}\,\left[-\frac{1}{2}\left(1-\alpha\right)\varrho+
      \frac{1}{2}\left(5-3\alpha\right)p\right].
\end{equation}
Again, setting $\alpha=1$ one arrives at the general relativistic
counterpart of the above equation
\[
  \left(\varrho+p\right)\,\frac{{\rm d}B}{{\rm d}r}=
  -\frac{{\rm d}p}{{\rm d}r}.
\]
Moreover, combining Eq.~(\ref{eq:rqrr}) and ~(\ref{eq:motion}) we obtain
the Rastall's counterpart of the Oppenheimer-Volkov equation (see
e.g.~\cite{schutz:1985,wald:1984})
\begin{equation}
  \label{eq:opvol}
\frac{{\rm d}{\cal P}}{{\rm d}r}=-
\left({\cal R}+{\cal P}\right)
\frac{\left({\cal M}+4\pi{\cal P}r^{3}/\alpha\right)}{r\left(r-2{\cal M}\right)}.
\end{equation}
Here, $p$ is always understood to be related to $\varrho$ by the equation
of state, which also determines the relation between ${\cal P}$ and ${\cal
  R}$. In full analogy to general relativity
equations~(\ref{eq:rqtt}),~(\ref{eq:rqrr}) and~(\ref{eq:opvol}) may be
collectively referred to as the equations of structure for spherical stars
in Rastall's generalization of the Einstein gravitational theory.

Let us look for the counterpart of the famous Schwarzschild solution. Let
$p,\varrho\neq0$ only for $r<R$, where $R$ corresponds to a surface of a
spherical star. Then Eq.~(\ref{eq:rqttint}) gives
\[
{\cal M}\left(R\right)=\frac{4\pi}{\alpha}\,\int\limits_{0}^{R}{\cal
  R}r^2{\rm d}r=\int\limits_{0}^{R}4\pi\varrho r^2{\rm
  d}r+\frac{3}{2}\,\frac{1-\alpha}{\alpha}\int\limits_{0}^{R}4\pi\left(\varrho-p\right)r^2\,{\rm d}r,
\]
where the first part corresponds to the standard Schwarzschild mass $M_{\rm
schw}$ and the second represents ``Rastall's'' correction. For a distant
observer at $r>R$ the value ${\cal M}={\cal M}\left(R\right)$ is constant
and the metric coefficients read as
\[
g_{rr}=\exp\left[2A\left(r,\alpha\right)\right]=\frac{1}{1-2{\cal M}/r},\qquad
g_{tt}=-\exp\left[2B\left(r,\alpha\right)\right]=-\left(1-\frac{2{\cal M}}{r}\right),\qquad
\]
where
\[
{\cal M}=M_{\rm schw}+\frac{3}{2}\,\frac{1-\alpha}{\alpha}\int\limits_{0}^{R}4\pi\left(\varrho-p\right)r^2\,{\rm d}r.
\]
Naturally, putting $\alpha=1$ we get the standard Schwarzschild
solution. Considering only dust ($p=0$) solution, which is theoretically
interesting, but physically highly non-realistic case, one come to the
result
\[
{\cal M}=M_{\rm
  schw}+\frac{3}{2}\,\frac{1-\alpha}{\alpha}\int\limits_{0}^{R}4\pi\varrho r^2\,{\rm d}r=
\frac{3-\alpha}{2\alpha}\,M_{\rm schw}.
\]
This puts a strong upper bound for the value of $\alpha$ -- evidently it must hold
that $\alpha<3$ to ensure ${\cal M}\geq0$.

The inconvenience of the value $\alpha=3$ can by illustrated by the
following example. Let us calculate the metric elements $A(r,\alpha=3)$ and
$B(r,\alpha=3)$ for an ideal perfect fluid with zero pressure $p=0$ (i.e.
for a dust) as a source.  In this case Rastall's
equations~(\ref{eq:rasteq}) turn out to be
\begin{equation}
  \label{eq:reqa3}
  G_{\mu\nu}=\frac{8\pi}{3}\,\left(T_{\mu\nu}-g_{\mu\nu}T\right)
\end{equation}
with $T=3p-\varrho=\varrho$ and functions ${\cal R}$, ${\cal P}$ listed in
the last row of Table~\ref{tab:alphpr}, particularly ${\cal R}=3p=0$ and
${\cal P}=2p+\varrho=\varrho$. Solving Eq.~(\ref{eq:rqtt}) we come to the
condition ${\rm d}{\cal M}/{\rm d}r=0$, which is identically valid for all
$r$, thus taking into account asymptotic flatness we get ${\cal M}=0$,
$A(r,\alpha=3)=\ln\left(1\right)/2=0$, $g_{rr}=\exp\left(2A\right)=1$,
which means the space-like part of the metric is flat.

On the other hand, Eq.~(\ref{eq:rqrr}) leads to
\[
\frac{{\rm d}B}{{\rm d}r}=\frac{4\pi}{3}\,\varrho r.
\]
Setting the conditions of asymptotic flatness $B(0)=B(\infty)=0$
again, one obtains
\[
\int\limits_{0}^{\infty} \frac{{\rm d}B}{{\rm d}r'}\,{\rm
  d}r'=B(\infty)-B(0)=0=
\int\limits_{0}^{r}\frac{4\pi}{3}\,\varrho r'\,{\rm
  d}r'+\int\limits_{r}^{\infty}\frac{4\pi}{3}\,\varrho r'\,{\rm d}r'=
\int\limits_{0}^{r}\frac{4\pi}{3}\,\varrho r'\,{\rm
  d}r'-B(r)
\]
and
\[
B(r)=\int\limits_{0}^{r}\frac{4\pi}{3}\,\varrho r'\,{\rm d}r',\qquad
g_{tt}=-\exp\left(2B\right)=-\exp\left(\int\limits_{0}^{r}\frac{4\pi}{3}\,\varrho
  r'\,{\rm d}r'\right).
\]
Thus the fluid affects only the $(t,t)$-part of the metric, while the
space remains Euclidean, which is physically acceptable only as an
approximation of weak gravitational fields.

\section{Machian property of Rastall's field equations}
\label{sec:mach}

An important consequence of Mach's principle is that inertial properties of
a mass body are determined by the distribution of mass-energy in the
neighbouring space.  Here, the problem arises how to adequately
characterize the physical state of this space.  It seems that the simplest
way to do it is either to take the gravitational potential in classical gravity,
or the metric in general relativity. That is, to suppose that the source of
gravitation, either mass-density or the elements of energy-momentum tensor,
depend either on the gravitational potential or on the corresponding
metric, respectively.

Let us recall Rastall's equations in the form~(\ref{eq:rasteq})
$$ G_{\mu\nu}=R_{\mu\nu}-\frac{1}{2}\,g_{\mu\nu}R =\frac{8\pi}{\alpha}\,\left(T_{\mu\nu}+
\frac{1-\alpha}{2}\,g_{\mu\nu}T\right)=\frac{8\pi}{\alpha}\left[T_{\mu\nu}+
a\left(\alpha\right)g_{\mu\nu}T\right],$$
where $a\left(\alpha\right)=\left(1-\alpha\right)/2$. We see that the
source contains elements of metric tensor.  This is in accord with the
conclusion of Mach's principle \cite{einstein:1956,nightingale:1977} which
says that the inert mass of a mass body depends on the magnitude and mutual
distance of the neighbouring mass distribution.  Thus, from the point of
view of Mach's principle, Rastall's field equations seem to be more
``Machian'' than those of the Einstein general relativity.

In the Newtonian gravity limit the metric in the Cartesian coordinates can
be put down form~\cite{hartle:2003,schutz:1985}
\[
  {\rm d}s^2=-\left(1+2\phi\right)\,{\rm d}t^2
  +\left(1-2\phi\right)\,\left({\rm d}x^2+{\rm d}y^2+{\rm d}z^2\right),
\]
where $\phi$ is the classical gravitational potential (typically,
$\phi=-M/r$ outside a spherical star of the mass $M$). Taking into account
the approximation of the energy momentum tensor
$$T_{tt}=\rho,\qquad T=-\rho$$
we derive from Eq.~(\ref{eq:rasteq})
\[
G_{tt}\approx
2\Delta\phi=\frac{8\pi}{\alpha}\,\left(T_{tt}+\frac{1-\alpha}{2}\,g_{tt}T\right)
=\frac{8\pi}{\alpha}\,\varrho\left[1+\frac{1-\alpha}{2}\left(1+2\phi\right)\right].
\]
Apparently, setting $\alpha=1$ we get the standard Poisson equation
$\Delta\phi=4\pi\varrho$, otherwise
\begin{equation}
  \label{eq:a}
  \Delta\phi(\vec r)=4\pi\varrho(\vec r)\left[A+B\phi(\vec r)\right],
\end{equation}
where
$$A=1+\frac{3(1-\alpha)}{2\alpha}\quad {\rm and}\quad
B=\frac{1-\alpha}{\alpha}\,\phi(\vec r).$$
Eq.~(\ref{eq:a}) shows that in the classical limit the source term on the
right-hand side of Rastall's equations depends on the potential of
gravitational field in a specific way, which puts a strong
constraint on $\alpha$.  It should be approximately equal to 1 in order to
obtain the familiar Poisson equation for weak gravitational
fields. However, for large mass concentrations and strong gravitational
fields Eq.~(\ref{eq:a}) might play a significant role in astrophysics and cosmology
\cite{alrawaf:1996pl}, even for small deviation of $\alpha$
from $1$.  The fact that the source term of Eq.~(\ref{eq:a})
depends on the potential $\phi(\vec r)$ might be understood as an effect
consistent with Mach's principle in non-relativity physics because the mass
density $\varrho$ depends on the all neighbouring mass-energy distribution
through the scalar potential $\phi(\vec r)$.  Moreover, Eq.~(\ref{eq:a}) as
well as Eq.~(\ref{eq:rasteq}) shows that for $\alpha \neq 1$ the source
terms depend, contrary to general relativity, on spacetime coordinates,
i.e. in Rastall's theory the concept of variable mass appears.

The concept of variable rest mass in the theory of gravity is not new.  It
appears in Dicke's reformulation of Brans-Dicke theory in which the metric
obeys Einstein's equations, but in which rest masses vary in particular way
\cite{dicke:1962} or in Bekenstein's theory of the universal scalar field
\cite{bekenstein:1977} to mention only two examples.  The introduction of
variable rest masses in the theory of gravitation is part of the general
program for the possible incorporating some features of Mach's principle into
general relativity which may be reached mainly by the following
assumptions: (i) that particles interact with some external fields as they
move on trajectories \cite{bekenstein:1977} or by
Mannheim's~\cite{mannheim:1993} assumption of an external scalar field, which
gives a position dependent coupling to motion; (ii) that particles move on
standard geodesics but the Christoffel symbols are modified because the
equation of motion for the metric is different from the standard one. This
can be reached either through a fundamental replacement of the Einstein
curvature tensor $G_{\mu\nu}$ by another equally covariant second-rank
tensor, or through introducing a more complicated energy-momentum tensor $T_{\mu\nu}$
than it is conventionally considered, keeping the common Einstein tensor at the
same time.

\section*{Concluding remark}
Rastall modified Einstein' equations by replacing the second term in the
Einstein tensor. The advantage of this substitution is that Rastall's
equations reduce to Einstein's one for empty space-time. Differences
between Einstein's and Rastall's equations arise when one investigates
gravitating systems with non-zero sources. We have showed that considerable
differences between both theories exists also regarding the incorporation
of Mach's principle in the theory of gravitation.  Hence, Rastall's
equations represent a set of gravity equations which includes also the
Einstein ones. In this sense, they can be considered as a generalization of
general relativity.


\end{document}